\documentclass[fleqn]{2023SCGE}
\usepackage{float}
\usepackage{multirow}
\begin{document}

\ensubject{subject}

\ArticleType{Article}
\Month{XXX}
\Vol{XXX}
\No{XXX}
\DOI{XXX}
\ArtNo{XXX}
\ReceiveDate{XXX}
\AcceptDate{XXX}
\OnlineDate{XXX}

\title{Reveal short range interactions between $u/d$ quarks in the $NN$, $D_{03}$, and $D_{30}$ systems}{Reveal short range interactions between $u/d$ quarks in the $NN$, $D_{03}$, and $D_{30}$ systems}

\author[1,2,3]{Qi-Fang L\"u}{lvqifang@hunnu.edu.cn}%
\author[4,5]{Yu-Bing Dong}{dongyb@ihep.ac.cn}
\author[4]{Peng-Nian Shen}{}
\author[4]{Zong-Ye Zhang}{}%

\AuthorMark{Qi-Fang L\"u}

\AuthorCitation{Qi-Fang L\"u, Yu-Bing Dong, Peng-Nian Shen, Zong-Ye Zhang}

\address[1]{Department of Physics, Hunan Normal University, Changsha 410081, China}
\address[2]{Key Laboratory of Low-Dimensional Quantum Structures and Quantum Control of Ministry of Education, Changsha 410081, China}
\address[3]{Key Laboratory for Matter Microstructure and Function of Hunan Province, Hunan Normal University, Changsha 410081, China}
\address[4]{Institute of High Energy Physics, Chinese Academy of Sciences, Beijing 100049, China}
\address[5]{School of Physical Sciences, University of Chinese Academy of Sciences, Beijing 101408, China}


\abstract{The dynamic mechanism of short range interactions between $u/d$ quarks remains an open and challenging problem. In order to reveal this quark dynamics, we perform a systematic analysis of $NN$, $D_{03}$, and $D_{30}$ systems in the (extended) chiral SU(3) constituent quark models. By comparing results calculated with different models and different parameter sets, the effects of one gluon exchange and vector meson exchange terms are carefully examined. The results indicate that the vector meson exchange interactions dominate the short range interactions between $u/d$ quarks, whereas the small residual one gluon exchange coupling strength is also allowed.}



\maketitle


\begin{multicols}{2}
\section{Introduction}\label{section1}
Understanding the strong interactions between constituent quarks is a fundamental and intriguing topic in hadron physics. It is now evident that the hadronic dynamics is governed by quantum chromodynamics (QCD), which exhibits asymptotic freedom in the short range and color confinement in the long distance. Owing to the intricacy of non-perturbative features, rigorous solutions for QCD in low-energy region become extremely challenging, and then one has to seek lattice QCD calculations, effective field theory, and various phenomenological models. In particular,  some constituent quark models~\cite{Isgur:1977ef,Shen:1997jd,Glozman:1995fu,Oka:1981ri,Oka:1981rj,Faessler:1983yd,Zhang:1997ny,Garcilazo:2001md,Dai:2003dz,Furuichi:2002gi,Ping:1998si,Capstick:1986ter,He:2023ucd} were proposed in spirit of 
\Authorfootnote

\noindent
QCD, which has gained considerable successes in describing the internal structures and properties of hadrons, such as baryon spectroscopy and nucleon-nucleon scattering. Despite these achievements, several long-standing problems and disputes exist in the constituent quark models~\cite{Isgur:1999jv,Meng:2023jqk,Pirjol:2008gd,Liu:1998um,Isgur:1999ic,Liu:1999kq}, especially for the short range interactions between light quarks. Resolving these difficulties is undoubtedly important for further understanding the strong interaction in the lower energy QCD region.

Historically, the study of the origin of strong interaction began with investigating the nuclear force on the hadron level, where the nucleon-nucleon scattering data were utilized as a checkpoint for judging theoretical models. The most successful model is so-called the one boson exchange model, where the pion and $\sigma$ meson  exchanges as well as the vector mesons $\rho$ and $\omega$ exchanges are also introduced to reproduce the wealth of experimental data. With a large number of coupling strengthens and cutoff parameters, the nucleon-nucleon interaction can be well described~\cite{Stoks:1994wp,Jade:1996hv,Long:2011xw,Lu:2021gsb,Machleidt:1987hj,Shen:2019dls,Wiringa:1994wb,Machleidt:2011zz,Epelbaum:2008ga}. The model on hadron level has also been extended to various hadronic systems, and one of the findings on studying these systems clearly demonstrated that vector meson exchange interactions are mainly responsible for the short range interactions and make certain contribution to the formation of some loosely bound molecular states~\cite{Guo:2017jvc,Chen:2022asf,Dong:2021bvy,Dong:2022cuw,Chen:2023def,Xie:2024wbd,Broniowski:1985kj,Meissner:1987ge,Kaiser:1998wa,Molina:2009ct,Wu:2010jy}.

Since the QCD was established, the research of the strong interaction has advanced to the quark level. One of the successful models is the constituent quark model, which allows one to explore the underlying quark dynamics and characterize as many hadronic systems as possible using a small number of unified parameters. It is found that the short range interaction between hadrons, in particular the repulsive core, is closely related to the one-gluon exchange and the quark exchange. Furthermore, on the quark level, it is easy to deal with the quark exchange effect and hidden color configurations that are absent on the hadron level. Specifically, different quark-quark ($q-q$) potentials are used in distinct constituent quark models. In the model we used, namely the chiral SU(3) or extended chiral SU(3) constituent quark model, the pesudoscalar and scalar chiral field induces $q-q$ potentials and the confining $q-q$ potential is mainly responsible for the medium and long range interactions, whereas the one gluon exchange (OGE) and/or vector meson exchange (VME) induced interactions dominate the short range interaction. An inevitable question arises: which interaction provides the short range interaction between $u/d$ quarks, OGE, VME, or both of interacting terms? Until now, the dynamic mechanism of short range interactions between $u/d$ quarks remains an open and challenging problem.

In fact, in the bound state problem of the $NN$ system, the roles of OGE and VME potentials in the nucleon-nucleon interaction have intensively been investigated, but there is still no agreement on who provides the short range interaction. One of the main reasons is that the studied deuteron is a loosely bound state of the $NN$ system, and the size of the deuteron is quite large. As a result, the two interacting nucleons are so far apart that they interact primarily through the long range interaction arising from pion meson exchange and are very insensitive to the short range interactions. To reveal the mechanism of short range interaction between $u/d$ quarks,  it is better to find a system that is not only more compact than the deuteron but also has experimental data about its properties. One example of such a system is the dibaryon $D_{03}$, where the notation $D_{IJ}$ stands for a dibaryon with isospin $I$ and spin $J$. In order to make the quark model more reliable and more applicable, it is also important to verify and adjust the model parameters as much as possible with the observed data of other dibaryons. Because of this, the mirror state $D_{30}$ of the dibaryon $D_{03}$ should also be studied, where the short range interaction of vector meson induced potentials differ entirely from those of $D_{03}$. Actually, there have been many research works on these two states and related topics in the literature both experimentally~\cite{Bashkanov:2008ih,Adlarson:2011bh,Clement:2016vnl,Adlarson:2014pxj,Adlarson:2014ozl,WASA-at-COSY:2016bha,Clement:2020mab} and theoretically~\cite{Dyson:1964xwa,Gal:2013dca,Gal:2014zia,Huang:2013nba,Chen:2014vha,Park:2015nha,Yuan:1999pg,Dai:2005kt,Huang:2014kja,Dong:2015cxa,Dong:2023xdi,Huang:2015nja,Lu:2017uey,Ping:2000dx,Bashkanov:2013cla,Kukulin:2022gze,Zhang:2022tzx,Dai:2023ofz,Bashkanov:2023yca,Li:1999dm,Zhang:2020dma,Zhang:2021vsf,Liu:2024ygk}. Furthermore, there still exist some dissenting opinions for the observed peak structure in experiments, where the authors claimed that the peak structure should be linked to a triangle singularity in the reaction~\cite{Ikeno:2021frl,Molina:2021bwp}. In any event, the experimental and theoretical advances have sparked a broad interest in the dibaryon research. Consequently, we would like to stress that the $D_{03}$ and $D_{30}$ systems offer excellent platforms for investigating the short range interactions between $u/d$ quarks at the quark level.

In this work, we perform a systematic analysis of $NN$, $D_{03}$, and $D_{30}$ systems in the (extended) chiral SU(3) constituent quark models. Firstly, for each mentioned quark model, we adjust the model parameters to get the best fit of the available data of the ground state masses of light baryons, binding energy of the deuteron, and the $NN$ phase shifts. Next, in terms of the same set of adjusted model parameters, we calculate the properties of $D_{03}$ and $D_{30}$ dibaryons. Finally, by comparing the results from various interaction models, namely models with distinct interactions, we study the effects of OGE and VME interactions.

This paper is organized as follows. In Sec.~\ref{sec:2}, the theoretical formalism of (extended) chiral SU(3) constituent quark model is introduced. We present the results and discussions of $NN$, $D_{03}$, and $D_{30}$ systems in Sec.~\ref{sec:3}. A summary is given in the last section.

\section{Method}\label{sec:2}

\subsection{Interactions}{\label{interaction}}

On the quark level, the interactions between constituent quarks are intermediated by the gluon and chiral fields. Here, we choose the chiral SU(3) quark model, where the interactive Lagrangian between the quark and chiral field can be expressed as
\begin{eqnarray}
	{\cal L}_I^{ch}&=&-g_{ch} \bar \psi \Bigg (\sum_{a=0}^8 \lambda_a \sigma_a +
	i \gamma_5 \sum_{a=0}^8 \lambda_a \pi_a \Bigg ) \psi,
\end{eqnarray}
where $g_{ch}$ is the coupling constant of quark with the chiral
field, $\psi$ is the quark field, and $\sigma_a$ and $\pi_a$
$(a=0,1,...,8)$ are the scalar and pseudo-scalar nonet chiral
fields, respectively. According to this Lagrangian, the corresponding Hamiltonian can be
obtained
\begin{eqnarray}
	{\cal H}_I^{ch}&=&g_{ch} F(\boldsymbol{q}^2) \bar \psi \Bigg (\sum_{a=0}^8
	\lambda_a \sigma_a +i \gamma_5 \sum_{a=0}^8 \lambda_a \pi_a \Bigg ) \psi.
\end{eqnarray}
Here, a form factor $F(\boldsymbol{q}^2)$ is introduced to describe the structures of the chiral fields, which is usually taken as~\cite{Dai:2003dz,Valcarce:2005em}
\begin{eqnarray}
	F(\boldsymbol{q}^2)=\Bigg (\frac{\Lambda^2}{\Lambda^2+\boldsymbol{q}^2} \Bigg )^{1/2}.
\end{eqnarray}
The $\Lambda$ is the cutoff mass, which corresponds to the scale of the chiral symmetry breaking. From this Hamiltonian, one can easily derive the quark-quark interaction $V^{\sigma_a}$ and $V^{\pi_a}$ arising from the chiral fields, which provide both medium and long range interactions. Besides the chiral field induced interactions, OGE potential $V^{OGE}$ and phenomenological confining potential $V^{conf}$ are still required, which correspond to the short range and long range interactions, respectively. Consequently, the total Hamiltonian for a six-quark system in the chiral SU(3) quark model can be given by
\begin{eqnarray}
	H=\sum_{i=1}^6 T_i -T_G + \sum_{j>i=1}^6 \Bigg (V_{ij}^{OGE}
	+V_{ij}^{conf}+V_{ij}^{ch} \Bigg ),
\end{eqnarray}
with
\begin{eqnarray}
	V_{ij}^{ch}=\sum_{a=0}^8 V_{ij}^{\sigma_a} + \sum_{a=0}^8 V_{ij}^{\pi_a},
\end{eqnarray}
where $T_i$ and $T_G$ are the kinetic energy operators for the
$i$-th quark and the center of mass motion, respectively. The $V_{ij}^{OGE}$, $V_{ij}^{conf}$, and $V_{ij}^{ch}$ stand for the OGE, confinement, and chiral field induced interactions between the
$i-$th and $j-$th quarks, respectively.

To better study the short-range interaction mechanism, the interactions between the quark and the vector meson fields are introduced in the extended chiral SU(3) quark model~\cite{Dai:2003dz}. The interactive Lagrangian is
\begin{eqnarray}
	{\cal L}_I^{\rm chv} = -g_{\rm chv} \bar{\psi}\gamma_\mu \lambda_a
	\rho^\mu_a \psi -\frac{f_{\rm chv}}{2M_N} \bar{\psi} \sigma_{\mu\nu}
	\lambda_a
	\partial^\mu \rho^\nu_a \psi.
\end{eqnarray}
Here the $\rho_{a}$ ($a=0,1,\cdots,8$) represent the vector nonet fields, and $g_{\rm chv}$ and $f_{\rm chv}$ stand for the coupling constants for vector and tensor terms between quark and vector fields, respectively. After adding the vector meson exchange interaction, the chiral fields induced effective interaction between the $i-$th quark and the $j-$th quark, $V^{\rm ch}$ reads
\begin{eqnarray}
	V_{ij}^{\rm ch} \,=\, \sum_{a=0}^8 V_{ij}^{\sigma_a} + \sum_{a=0}^8 V_{ij}^{\pi_a} + \sum_{a=0}^8 V_{ij}^{\rho_a},
\end{eqnarray}
with $V^{\rho_a}$ being the quark-quark interaction potential induced by vector-meson exchanges. The vector-meson exchange potential $V^{\rho_a}$ is also short range, which competes with the OGE interaction. Therefore, it is more suitable to investigate the short range interaction mechanism in the extended chiral SU(3) quark model. Moreover, it should be noted that the potentials induced by chiral fields have entirely different operators in the spin-flavor-color space compared with the OGE and confining potentials, and they coexist and work in tandem, that is, complementing one another in the (extended) chiral SU(3) quark models. The explicit expressions of these potentials can be found in Refs.~\cite{Zhang:1997ny,Dai:2003dz,Huang:2015nja}.

\subsection{Parameters}{\label{Para}}

In our calculation, the $\eta$ and $\eta'$ mesons are mixed by
$\eta_1$ and $\eta_8$, and the mixing angle $\theta_{\eta}$ is taken
to be the usual value with $\theta_{\eta}=-23^\circ$. For the $\omega$ and
$\phi$ mesons, we adopt the flavor wave functions $(u\overline{u}+d\overline{d})/\sqrt{2}$ and
$s\overline{s}$ respectively, that is, they are mixed by $\omega_1$
and $\omega_8$ with the ideal mixing angle $\theta_{\omega}=-54.7^\circ$. In the non-strange multi-quark systems, the $u$ or $d$ quark mass is taken to be $m_{u/d}=313~\rm{MeV}$. The harmonic-oscillator width parameter $b_u$ in the Gaussian wave function for each $u$ or $d$ quark is chosen to be around $0.45~\rm{fm}$, and the effects of its variation will be discussed in the following section.

All meson masses are taken from the experimental values except for the $\sigma$ meson. According to the dynamical vacuum spontaneous breaking mechanism, its value should fulfill \cite{Dai:2003dz}
\begin{eqnarray}
	m_{\sigma}^2= (2 m_u)^2 + m_{\pi}^2.
\end{eqnarray}
This gives us a strong constraint for the mass of $\sigma$ meson. As in previous calculations, we treat it as an adjustable parameter by fitting the binding energy of deuteron, which falls within the reasonable range around $500\sim700$ MeV. The parameter $\Lambda$ is expected to be around 1 GeV, similar to that in one boson exchange model on the hadron level. The exact value depends on the phenomenological model and is finally determined by the experimental data. To reduce the numbers of free parameters, in our model we adopt an effective value of 1100 MeV for $\Lambda$, which is also close to the chiral symmetry breaking scale.

In the chiral SU(3) quark model, the coupling constant between the quark field and the scalar and pseudo-scalar chiral fields $g_{\rm ch}$ is determined according to the relation
\begin{eqnarray}
	\frac{g^{2}_{\rm ch}}{4\pi} = \left( \frac{3}{5} \right)^{2}
	\frac{g^{2}_{NN\pi}}{4\pi} \frac{m^{2}_{u}}{M^{2}_{N}},
\end{eqnarray}
with $g^{2}_{NN\pi}/4\pi=13.67$. After the parameters of chiral fields are fixed, the coupling constant $g_u$ of OGE interaction can be determined by the mass gap of $\Delta-N$. The masses of ground states $N$ and $\Delta$ with 939 and 1232 MeV, respectively, are employed here. Then, the confinement strength $a_{uu}^c$ and zero-point energy $a_{uu}^{c0}$ are fixed by the stability condition and the mass of the nucleon, respectively.

In the extended chiral SU(3) quark model, the VME interactions are also added, and three kinds of coupling constants are taken into consideration here. Two kinds of coupling constants are identical to those in previous works~\cite{Dai:2003dz,Huang:2015nja}: $g_{\rm chv}=2.351$, $f_{\rm chv}/g_{\rm chv}=0$, and $g_{\rm chv}=1.972$, $f_{\rm chv}/g_{\rm chv}=2/3$. In the present calculation, to study the mechanism of the short range interaction, we additionally consider an extreme scenario where OGE interaction is excluded and only VME interactions are responsible for short range interaction. In this situation, the coupling constant $g_u$ of the OGE interaction is forced to equal zero, where the OGE term is replaced by the VME interactions. Then the $g_{\rm chv}$ is no longer a free parameter, which is completely determined by the constraint of the $\Delta-N$ mass gap. The typical value of  $g_{\rm chv}$ in this case is 2.536 when $b_u$ equals to $0.45~{\rm fm}$.

All the parameters used in the present work are tabulated in Table~\ref{para}. The Sets I, II, and III represent three cases with $b_u=0.43$, 0.45, and 0.47 fm, respectively, because the $b_u$ is the main source of uncertainties of (extended) chiral SU(3) quark models. Here, we vary the parameter $b_u$ to examine its dependence for the results, which can be regarded as an effort to discuss the uncertainties. It should be mentioned that the numbers of coupling constants in our model are greatly reduced by the constraints of the necessary SU(3) chiral symmetry in nuclear dynamics, and the parameters, including cutoff values, are also extracted and determined from the experimental data. Since the coupling constant $g_{\pi N N}$ and masses of $\Delta$, $N$, and deuteron are measured and determined rather precisely in experiments, the uncertainties resulting from coupling constant $g_{ch}$, mass of $\sigma$ meson $m_\sigma$, coupling constant $g_u$, confinement strength $a^c_{uu}$, and zero-point energy $a^{c0}_{uu}$ are tiny and do not affect all the results and conclusions significantly. For instance, the variation of $m_\sigma$ about 1 MeV will result in the change of the binding energy of deuteron about 0.2 MeV.

When all the parameters in the potentials are determined, two-baryon systems on the quark level can be studied by using the resonating group method (RGM). Then, one can dynamically obtain the partial wave phase shifts of $NN$ scattering and properties for dibaryons. More details about the (extended) chiral SU(3) quark model and RGM can be found in previous works~\cite{Zhang:1997ny,Dai:2003dz,Huang:2015nja}.

\begin{table}[H]
	\footnotesize
	\begin{threeparttable}
		\caption{\label{para} Model parameters. The meson masses and the cutoff mass are $m_{\sigma'}=980$ MeV, $m_{\epsilon}=980$ MeV, $m_{\pi}=138$ MeV, $m_{\eta}=548$ MeV, $m_{\eta'}=958$ MeV, $m_{\rho}=770$ MeV, $m_{\omega}=782$ MeV, and $\Lambda=1100$ MeV.}
		\doublerulesep 0.1pt \tabcolsep 9pt 
		\begin{tabular}{cccccc}
			\toprule
			\multirow{2}{*}{Set I} &\multirow{2}{*}{SU(3)} & \multicolumn{3}{c}{Ext. SU(3)}  \\ \cline{3-5}
			&	  & f/g=0 & f/g=2/3 & no OGE \\
			\hline
			$b_u$ (fm)  & 0.43 & 0.43 & 0.43 & 0.43\\
			$g_{\rm ch}$    & 2.621 & 2.621 & 2.621 & 2.621 \\
			$g_{\rm chv}$   & $\cdot\cdot\cdot$ & 2.351 & 1.972 & 2.358  \\
			$m_\sigma$ (MeV) & 702 & 549 & 561 & 547 \\
			$g_u$     & 0.672 & 0.052 & 0.269 &  $\cdot\cdot\cdot$ \\
			$a^c_{uu}$ (MeV/fm$^2$) & 117.47 & 58.23 & 52.48 & 57.84\\
			$a^{c0}_{uu}$ (MeV)  & $-113.80$ & $-86.75$ & $-77.57$ &  $-86.52$\\
			\hline\hline
			\multirow{2}{*}{Set II} &\multirow{2}{*}{SU(3)} & \multicolumn{3}{c}{Ext. SU(3)}  \\ \cline{3-5}
			&	  & f/g=0 & f/g=2/3 & no OGE \\
			\hline
			$b_u$ (fm)  & 0.45 & 0.45 & 0.45 & 0.45\\
			$g_{\rm ch}$    & 2.621 & 2.621 & 2.621 & 2.621 \\
			$g_{\rm chv}$   & $\cdot\cdot\cdot$ & 2.351 & 1.972 & 2.536  \\
			$m_\sigma$ (MeV) & 668 & 535 & 547 & 522 \\
			$g_u$     & 0.730 & 0.273 & 0.385 &  $\cdot\cdot\cdot$ \\
			$a^c_{uu}$ (MeV/fm$^2$) & 92.39 & 48.43 & 43.33 & 41.46\\
			$a^{c0}_{uu}$ (MeV)  & $-92.61$ & $-75.46$ & $-66.38$ &  $-72.95$\\
			\hline\hline
			\multirow{2}{*}{Set III} &\multirow{2}{*}{SU(3)} & \multicolumn{3}{c}{Ext. SU(3)}  \\ \cline{3-5}
			&	  & f/g=0 & f/g=2/3 & no OGE \\
			\hline
			$b_u$ (fm)  & 0.47 & 0.47 & 0.47 & 0.47\\
			$g_{\rm ch}$    & 2.621 & 2.621 & 2.621 & 2.621 \\
			$g_{\rm chv}$   & $\cdot\cdot\cdot$ & 2.351 & 1.972 & 2.715  \\
			$m_\sigma$ (MeV) & 637 & 524 & 533 & 498 \\
			$g_u$     & 0.788 & 0.394 & 0.483 &  $\cdot\cdot\cdot$ \\
			$a^c_{uu}$ (MeV/fm$^2$) & 72.16 & 39.79 & 35.24 & 29.22\\
			$a^{c0}_{uu}$ (MeV)  & $-72.77$ & $-64.00$ & $-54.84$ &  $-61.38$\\
			\bottomrule
		\end{tabular}
	\end{threeparttable}
\end{table}

\section{Results and discussions}\label{sec:3}
With the above potentials and parameters, we can calculate the properties of deuteron, $NN$ scattering, $D_{03}$, and $D_{30}$ to study the short range interactions between $u/d$ quarks. The binding energy, root-mean-squared (RMS) radius~\footnote{The RMS radius of a six-quark system with a single-cluster structure, denoted as $\mathcal{R}_{6q}$, is usually calculated by~\cite{Dai:2023ofz,Zhang:2006dy}
	\begin{eqnarray}
		{\mathcal R}_{6q}=\sqrt{\frac{1}{6}\sum_{i=1}^6\left \langle ( \boldsymbol{r}_i-\boldsymbol{R}_{\rm c.m.})^2\right \rangle}. \nonumber
	\end{eqnarray}
where $\boldsymbol{r}_i$ and $\boldsymbol{R}_{\rm c.m.}$ stand for the coordinates of the $i^{th}$ quark and the c.m. of the six-quark system, respectively. However, if the six-quark system clearly exhibits a two-cluster structure, especially a quasi-molecular structure like the deuteron, its RMS radius could be expressed as ${\mathcal R}_{AB}$, where subscripts $A$ and $B$ stand for the $A$ and $B$ clusters, respectively, and would be evaluated by
		\begin{eqnarray}
			{\mathcal R}_{AB}=\sqrt{ \langle {\boldsymbol{r}_{AB}}^2 \rangle}, \nonumber
		\end{eqnarray}
where $\boldsymbol{r}_{AB}$ is the relative coordinate between $A$ and $B$ clusters. It should be noted that the ${\mathcal R}_{AB}$ and ${\mathcal R}_{6q}$ for a definite system usually have different expected values due to different operators in the above formulas, where the different inner structure and shape are assumed.}, and fraction of channel wave functions for deuteron are listed in Table~\ref{deuteron}. The calculated binding energy for deuteron is limited to $2.10\sim2.30~\rm{MeV}$, which fixes the model parameters, such as the mass of $\sigma$ meson. The resultant RMS radii and the percentages of partial waves in different models are almost the same. Specifically, the resultant percentages for the $S-$wave and $D-$wave in the deuteron are $93\sim95\%$ and $5\sim7\%$, respectively. This indicates that all the chiral SU(3) and extended chiral SU(3) constituent quark models used here can well describe the deuteron. It should be mentioned that in Table~\ref{deuteron}, the fact that the $\mathcal{R}_{NN}$ value, which agrees with the experimental data, is much greater than $\mathcal{R}_{6q}$ indicates that deuteron is not a compact six-quark state but a quasi-molecular state. The RMS radius ${\mathcal{R}}_{NN}$ of $2.4\sim2.5~{\rm fm}$ shows a large separation between constituent nucleon clusters. In other words, the proton and neutron in deuteron feel each other mainly through the long range interactions and are insensitive to the short range interactions. Consequently, the deuteron is not an ideal platform to distinguish the dynamic mechanism of short range interactions.

\begin{table}[H]
	\footnotesize
	\begin{threeparttable}\caption{\label{deuteron} Binding energy, RMS radius, and fraction of channel wave functions for deuteron in different models and parameters. The units for binding energy, RMS, and fraction are in MeV, fm, and percentage, respectively.}
		\doublerulesep 0.1pt \tabcolsep 9pt 
		\begin{tabular}{cccccc}
			\toprule
		\multirow{2}{*}{Set I} &\multirow{2}{*}{SU(3)} & \multicolumn{3}{c}{Ext. SU(3)}  \\ \cline{3-5}
		&	  & f/g=0 & f/g=2/3 & no OGE \\
		\hline
		Binding energy  & 2.25 & 2.20 & 2.22 & 2.24\\
			${\mathcal R}_{NN}$   & $ 2.43 $ & $2.48 $ & $ 2.47 $ & $2.47 $  \\ 
		${\mathcal R}_{6q}$   & 1.31 & 1.33 & 1.33 & 1.33 \\
		Fraction of $(NN)_{L=0}$  & 92.88 & 94.47 & 94.52 & 94.48  \\
		Fraction of $(NN)_{L=2}$ & 7.12 & 5.53 & 5.48 & 5.52 \\
		\hline\hline
		\multirow{2}{*}{Set II} &\multirow{2}{*}{SU(3)} & \multicolumn{3}{c}{Ext. SU(3)}  \\ \cline{3-5}
		&	  & f/g=0 & f/g=2/3 & no OGE \\
		\hline
		Binding energy  & 2.27 & 2.25 & 2.20 & 2.17\\
			${\mathcal R}_{NN}$   & $ 2.45 $ & $ 2.49 $ & $ 2.49 $ & $ 2.50 $  \\ 
		${\mathcal R}_{6q}$    & 1.33 & 1.34 & 1.35 & 1.35 \\
		Fraction of $(NN)_{L=0}$   & 93.09 & 94.65 & 94.70 & 94.90  \\
		Fraction of $(NN)_{L=2}$ & 6.91 & 5.35 & 5.30 & 5.10 \\
		\hline\hline
		\multirow{2}{*}{Set III} &\multirow{2}{*}{SU(3)} & \multicolumn{3}{c}{Ext. SU(3)}  \\ \cline{3-5}
		&	  & f/g=0 & f/g=2/3 & no OGE \\
		\hline
		Binding energy  & 2.27 & 2.13 & 2.24 & 2.22\\
			${\mathcal R}_{NN}$   & $  2.46 $ & $  2.51 $ & $ 2.51 $ & $  2.52 $  \\ 
		${\mathcal R}_{6q}$    & 1.34 & 1.36 & 1.36 & 1.37 \\
		Fraction of $(NN)_{L=0}$   & 93.30 & 94.84 & 94.87 & 95.32  \\
		Fraction of $(NN)_{L=2}$ & 6.70 & 5.16 & 5.13 & 4.68 \\
			\bottomrule
		\end{tabular}
	\end{threeparttable}
\end{table}

\begin{figure}[H]
	\centering
	\includegraphics[scale=0.46]{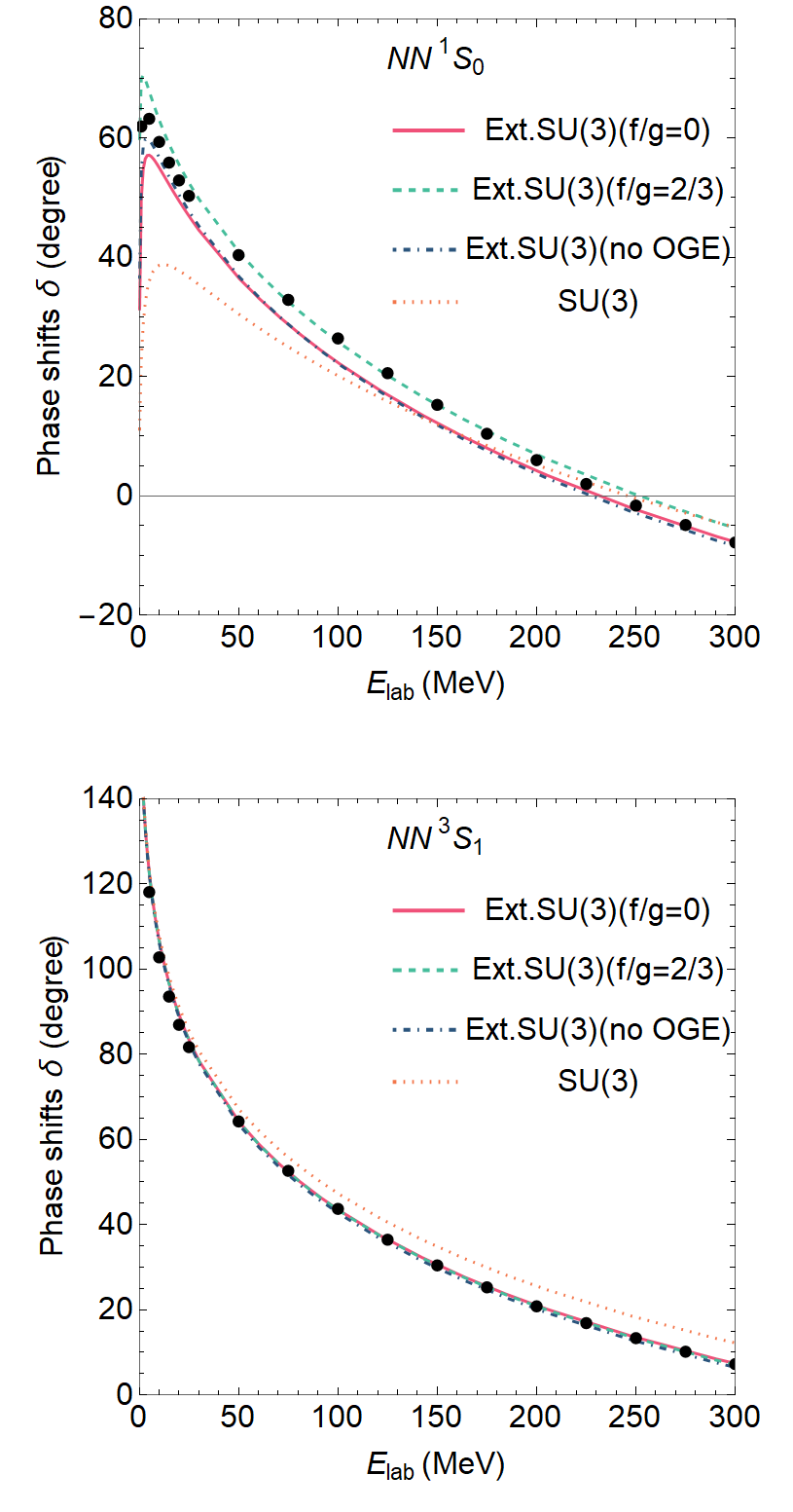}
   \caption{The $S-$wave phase shifts of $NN$ scattering in different quark models with $b_u=0.45~\rm{fm}$. The black points stand for the experimental data~\cite{Arndt:1994br}.}
	\label{phaseshift}
\end{figure}

The $S-$wave phase shifts of $NN$ scattering in different quark models with $b_u=0.45~\rm{fm}$ are displayed in Figure~\ref{phaseshift}. From this figure, one sees that resultant $^1S_0$ phase shifts by using the chiral SU(3) constituent quark model underestimate the experimental data, while the results from the extended chiral SU(3) constituent quark model with varied $g_{\rm chv}$ are obviously more reasonable. Meanwhile, the obtained $^3S_1$ phase shifts in either the chiral SU(3) constituent quark model or the extended chiral SU(3) constituent quark model can provide an excellent description of the experimental data. Furthermore, changes in $b_u$ around 0.45 fm have almost no effect on the behaviors of these phase shifts. Here, we would like to highlight that when the VME potential is introduced, namely in the extended chiral SU(3) constituent quark model, there are two key aspects: (1) the data of the $NN$ phase shifts, in particular the $^1S_0$ phase shifts, can naturally be reproduced; (2) the strength of OGE interaction is greatly reduced.

Since the deuteron is a loosely bound state and insensitive to the short range interactions, we move on to other nonstrange dibaryons, especially the more compact ones. Clearly, $d^*(2380)$, which is usually regarded as a dibaryon $D_{03}$ composed of $\Delta \Delta$ and hidden-color components (CC) in the literature, satisfies this demand. The results for $D_{03}$ and its mirror state $D_{30}$ are listed in Table~\ref{dibaryons}. From this table, one sees that the binding energy of $D_{03}$ is about $19\sim 32$ MeV, which is significantly different from the measured data for $d^*(2380)$. Actually, one cannot obtain a reasonable binding energy by using the chiral SU(3) quark model even with $b_u=0.5~\rm{fm}$ and large $g_u$~\cite{Huang:2015nja,Yuan:1999pg}. In Ref.~\cite{Yuan:1999pg}, $b_u=0.6$ fm is adopted to predict the reasonable binding energy for $d^*(2380)$. However, with this $b_u=0.6$ fm, the coupling strengthen $g_u$ of the OGE term will be significantly large, namely, it will be greater than 1, which seems unnatural. Meanwhile, in the case of the extended chiral SU(3) constituent quark model with $b_u=0.43 \sim 0.47~\rm{fm}$ and $f/g=0$, the resultant binding energy of $d^*(2380)$ is about $73\sim87$ MeV, which is compatible with the experimental data of $84~{\rm MeV}$. However, when the ratio of $f/g$ shifts from 0 to 2/3, the obtained binding energy of $d^*(2380)$ is about $58\sim75$ MeV, which underestimates the data. The reason may be due to the introduction of the tensor component in the vector-chiral-field-induced $q-q$ interaction; in order to better explain the available $NN$ data, the vector meson coupling constant $g_{chv}$ has to be decreased. Moreover, if we further eliminate OGE completely, the binding energy of $D_{03}$ will be about $73\sim 111$ MeV, which is sensitive to the parameter $b_u$. In a word, the trend of change in the binding energy of $D_{03}$ with respect to the model modification is the same as that of deuteron. From these results, we believe that the short-range attractive feature of VME plays an important role in the formation of $D_{03}$, which is the same as that provided by OGE but is much stronger. However, in order to reduce the dependence of the binding energy on model parameters, say the width parameter $b_u$, it seems that it is necessary to have a certain amount of OGE interaction. In fact, in this way, not only the dependence of the binding energy on $b_u$ is reduced, but also the gluon coupling constant $g_u$ becomes smaller, such that $g_u$ is compatible with the QCD spirit. It is evident that these findings contradict the views in some literature~\cite{Oka:1981ri,Oka:1981rj} that the OGE is the only mechanism for the short range interaction for nonstrange dibaryons.

\begin{table*}[!htbp]
	\footnotesize
	\begin{center}
		\caption{\label{dibaryons} Binding energies, RMS radii, and fractions of channel wave functions for dibaryons $D_{03}$ and $D_{30}$ in different models and parameters. The units for binding energy, RMS, and fraction are in MeV, fm, and percentage, respectively.}
				\renewcommand{\arraystretch}{1.1}
			\doublerulesep 0.1pt
		\begin{tabular*}{18cm}{@{\extracolsep{\fill}}p{3.0cm}<{\centering}p{1.4cm}<{\centering}p{1.4cm}<{\centering}p{1.4cm}<{\centering}p{1.4cm}<{\centering}p{1.4cm}<{\centering}p{1.4cm}<{\centering}p{1.4cm}<{\centering}p{1.4cm}<{\centering}}
			\toprule
			\multirow{3}{*}{Set I} &  \multicolumn{4}{c}{$D_{03}$}  &  \multicolumn{4}{c}{$D_{30}$}  \\ \cline{2-5} \cline{6-9}
			&\multirow{2}{*}{SU(3)}  & Ext. SU(3) (f/g=0) & Ext. SU(3) (f/g=2/3) & Ext. SU(3) (no OGE)  &\multirow{2}{*}{SU(3)}  & Ext. SU(3) (f/g=0) & Ext. SU(3) (f/g=2/3) & Ext. SU(3) (no OGE)\\
			\hline
			Binding energy  & 18.54 & 72.84 & 58.27 & 73.39  & 10.92 & 4.96 & 10.71 & 5.00 \\
		${\mathcal R}_{\Delta \Delta/{\rm CC}}$  & $ 1.16 $ & $ 0.74 $ & $  0.79 $ & $ 0.73 $  &  $ 1.55 $ & $ 2.02 $ & $ 1.73 $ & $ 2.02 $ \\
			${\mathcal R}_{6q}$    & 0.98 & 0.76 & 0.79 & 0.76  &  1.10  & 1.25 & 1.14 & 1.25 \\
	
			Fraction of $({\Delta \Delta})_{L=0}$   & 46.52 & 33.68 & 35.69 & 33.63 & 64.40 & 77.96 & 71.94& 77.97 \\
			Fraction of $({\Delta \Delta})_{L=2}$ & 2.18 & 0.64 & 0.76 & 0.63 & $\cdot\cdot\cdot$& $\cdot\cdot\cdot$&$\cdot\cdot\cdot$ &$\cdot\cdot\cdot$ \\
			Fraction of $({\rm CC})_{L=0}$   & 51.30 & 65.68 & 63.54 & 65.74 & 35.60& 22.04& 28.06& 22.03  \\
			Fraction of $({\rm CC})_{L=2}$ & 0.00 & 0.00 & 0.00 & 0.00& $\cdot\cdot\cdot$& $\cdot\cdot\cdot$&$\cdot\cdot\cdot$ &$\cdot\cdot\cdot$ \\
			\hline\hline
			\multirow{3}{*}{Set II} &  \multicolumn{4}{c}{$D_{03}$}  &  \multicolumn{4}{c}{$D_{30}$}  \\ \cline{2-5} \cline{6-9}
			&\multirow{2}{*}{SU(3)}  & Ext. SU(3) (f/g=0) & Ext. SU(3) (f/g=2/3) & Ext. SU(3) (no OGE)  &\multirow{2}{*}{SU(3)}  & Ext. SU(3) (f/g=0) & Ext. SU(3) (f/g=2/3) & Ext. SU(3) (no OGE)\\
			\hline
			Binding energy  &25.09 & 80.08 & 66.40 & 91.21  & 12.30 & 6.00 & 11.65 & 5.28 \\
				${\mathcal R}_{\Delta \Delta/{\rm CC}}$  & $ 1.04 $ & $ 0.73 $ & $ 0.77 $ & $ 0.71 $  &  $ 1.50 $ & $ 1.94 $ & $ 1.67 $ & $ 2.01 $ \\
			${\mathcal R}_{6q}$    & 0.94 & 0.77 & 0.79 & 0.76  &  1.10  & 1.25 & 1.14 & 1.27 \\
			Fraction of $({\Delta \Delta})_{L=0}$   & 42.07 & 31.91 & 33.41 & 30.90 & 61.29 & 74.30 & 68.31& 76.19 \\
			Fraction of $({\Delta \Delta})_{L=2}$ & 1.59 & 0.49 & 0.57 & 0.41 & $\cdot\cdot\cdot$& $\cdot\cdot\cdot$&$\cdot\cdot\cdot$ &$\cdot\cdot\cdot$ \\
			Fraction of $({\rm CC})_{L=0}$   & 56.34 & 67.60 & 66.02 & 68.68 & 38.71& 25.70& 31.69& 23.81  \\
			Fraction of $({\rm CC})_{L=2}$ & 0.00 & 0.00 & 0.00 & 0.00& $\cdot\cdot\cdot$& $\cdot\cdot\cdot$&$\cdot\cdot\cdot$ &$\cdot\cdot\cdot$ \\
			\hline\hline
			\multirow{3}{*}{Set III} &  \multicolumn{4}{c}{$D_{03}$}  &  \multicolumn{4}{c}{$D_{30}$}  \\ \cline{2-5} \cline{6-9}
			&\multirow{2}{*}{SU(3)}  & Ext. SU(3) (f/g=0) & Ext. SU(3) (f/g=2/3) & Ext. SU(3) (no OGE)  &\multirow{2}{*}{SU(3)}  & Ext. SU(3) (f/g=0) & Ext. SU(3) (f/g=2/3) & Ext. SU(3) (no OGE)\\
			\hline
			Binding energy  &32.35 & 87.01 & 75.02 & 110.95  & 13.38 & 6.71 & 12.49 & 5.54 \\
				${\mathcal R}_{\Delta \Delta/{\rm CC}}$  & $ 0.95 $ & $ 0.73 $ & $  0.76 $ & $ 0.70 $  &  $ 1.46 $ & $  1.88 $ & $ 1.62 $ & $ 2.00 $ \\
            ${\mathcal R}_{6q}$    & 0.91 & 0.79 & 0.80 & 0.77  &  1.10  & 1.25 & 1.15 & 1.29 \\
			Fraction of $({\Delta \Delta})_{L=0}$   & 38.39 & 30.33 & 31.40 & 28.63 & 58.56 & 71.08 & 64.98& 74.51 \\
			Fraction of $({\Delta \Delta})_{L=2}$ & 1.15 & 0.38 & 0.42 & 0.27 & $\cdot\cdot\cdot$& $\cdot\cdot\cdot$&$\cdot\cdot\cdot$ &$\cdot\cdot\cdot$ \\
			Fraction of $({\rm CC})_{L=0}$   & 60.45 & 69.29 & 68.18 & 71.10 & 41.44& 28.92& 35.02& 25.49  \\
			Fraction of $({\rm CC})_{L=2}$ & 0.01 & 0.00 & 0.00 & 0.00& $\cdot\cdot\cdot$& $\cdot\cdot\cdot$&$\cdot\cdot\cdot$ &$\cdot\cdot\cdot$ \\
				\bottomrule
		\end{tabular*}
	\end{center}
\end{table*}

\begin{table*}[htbp]
	\footnotesize
	\begin{center}
		\caption{\label{sfc} Relevant SFC coefficients of central potentials for $D_{03}$ and $D_{30}$ systems. When the SFC coefficients are different between these two systems, the values for $D_{30}$ systems are listed in the parenthesis.}
		\doublerulesep 0.1pt
		\renewcommand{\arraystretch}{1.1}
		\begin{tabular*}{18cm}{@{\extracolsep{\fill}}m{1.7cm}<{\centering}m{0.6cm}<{\centering}m{1.48cm}<{\centering}m{1.48cm}<{\centering}m{1.48cm}<{\centering}m{1.7cm}<{\centering}m{0.6cm}<{\centering}m{1.48cm}<{\centering}m{1.48cm}<{\centering}m{1.48cm}<{\centering}}
			\toprule
			\multicolumn{2}{c}{$\hat{O}_{ij}$}& $\Delta\Delta / \Delta\Delta$ & $\Delta\Delta /{\rm CC}$ & ${\rm CC/ CC}$ &
			\multicolumn{2}{c}{$\hat{O}_{ij}$}& $\Delta\Delta / \Delta\Delta$ & $\Delta\Delta / {\rm CC}$ & ${\rm CC/ CC}$  \\ \hline
			1 & & 1 & 0 & 1 & $P_{36}$ & & $-1/9$ & $-4/9$ & $-7/9$\\ \hline
			\multirow{7}{*}{$\lambda^c_i \cdot \lambda^c_j$} & $\hat{O}_{12}$ & $-8/3$ & 0 & $-2/3$ &
			\multirow{7}{*}{$\sum\limits_{k=1}^3\lambda^F_i(k)\lambda^F_j(k)$} & $\hat{O}_{12}$ & 1 & 0 & $-1(1)$\\ \cline{2-5}\cline{7-10}
			& $\hat{O}_{36}$ & 0 & 0 & $-4/3$ &
			& $\hat{O}_{36}$ & $-5/3(1)$ & 0 & $-1/3(1)$\\  \cline{2-5}\cline{7-10}
			& $\hat{O}_{12}P_{36}$& 8/27 &  32/27  & 2/27  &
			& $\hat{O}_{12}P_{36}$ & $-1/9$ & $-4/9$ & $11/9(-7/9)$\\ \cline{2-5}\cline{7-10}
			& $\hat{O}_{13}P_{36}$& 8/27 &  32/27  & 20/27  &
			& $\hat{O}_{13}P_{36}$ & $-1/9$ & $-4/9$ & $5/9(-7/9)$ \\ \cline{2-5}\cline{7-10}
			& $\hat{O}_{16}P_{36}$ & 8/27 & $-4/27$ & 20/27 &
			&
			$\hat{O}_{16}P_{36}$ & $-1/9$ & $8/9(-4/9)$ & $5/9(-7/9)$\\ \cline{2-5}\cline{7-10}
			& $\hat{O}_{14}P_{36}$ & $-4/27$ & 2/27 & 35/27 &
			& $\hat{O}_{14}P_{36}$ & $1/3(-1/9)$ & $2/3(-4/9)$ & $0(-7/9)$ \\ \cline{2-5}\cline{7-10}
			& $\hat{O}_{36}P_{36}$ & $-16/27$ & 8/27 & 32/27 &
			& $\hat{O}_{36}P_{36}$ & $7/9(-1/9)$ & $4/9(-4/9)$ & $1/9(-7/9)$ \\  \hline\hline
			\multirow{7}{*}{$(\bm{\sigma}_i \cdot \bm{\sigma}_j)(\lambda^c_i \cdot \lambda^c_j)$} & $\hat{O}_{12}$ &$-8/3$ & 0 & $-2/3(-10/3)$ &
			\multirow{7}{*}{\shortstack{$(\bm{\sigma}_i \cdot \bm{\sigma}_j)$\\ $\sum\limits_{k=1}^3\lambda^F_i(k)\lambda^F_j(k)$}}& $\hat{O}_{12}$ & 1 & 0 & $-1$ \\ \cline{2-5}\cline{7-10}
			& $\hat{O}_{36}$ & 0 & $0(-16/9)$ & $-4/3(-20/9)$ &
			& $\hat{O}_{36}$ & $-5/3$ & 0 & $-1/3$ \\  \cline{2-5}\cline{7-10}
			
			& $\hat{O}_{12}P_{36}$ & 8/27 &  32/27  & 2/27(74/27)  & 
			& $\hat{O}_{12}P_{36}$ & $-1/9$ & $-4/9$ & 11/9\\ \cline{2-5}\cline{7-10}
			& $\hat{O}_{13}P_{36}$& 8/27 &  32/27  & 20/27(68/27)  &
			& $\hat{O}_{13}P_{36}$ & $-1/9$ & $-4/9$ & 5/9 \\ \cline{2-5}\cline{7-10}
			& $\hat{O}_{16}P_{36}$ & 8/27 & $-4/27(44/27)$ & 20/27(68/27) &
			&
			$\hat{O}_{16}P_{36}$ & $-1/9$ & 8/9 & 5/9\\ \cline{2-5}\cline{7-10}
			& $\hat{O}_{14}P_{36}$ & $-4/27(4/9)$ & 2/27(14/9) & 35/27(7/3) &
			& $\hat{O}_{14}P_{36}$ & 1/3 & 2/3 & 0 \\ \cline{2-5}\cline{7-10}
			& $\hat{O}_{36}P_{36}$ & $-16/27(112/27)$ & $8/27(-8/27)$ & 32/27(88/27) &
			& $\hat{O}_{36}P_{36}$ & 7/9 & 4/9 & 1/9 \\  
			\bottomrule
		\end{tabular*}
	\end{center}
\end{table*}

It should be specially mentioned that unlike the deuteron, the $D_{03}$ system is a deeply bound state with a large CC component, and the calculated RMS radius $\mathcal{R}_{\Delta\Delta/{\rm CC}}$ is about $0.73~\rm fm$. This compact structure arises from the large quark exchange effect and attractive short range interaction, and as a consequence, a very large hidden color component (CC) appears. The large quark exchange effect can be seen from spin-flavor-color (SFC) coefficients due to symmetry, which can be characterized by the averaged value of antisymmetrizer $\langle \mathcal{A}^{sfc} \rangle$ in the spin-flavor-color space. In the six identical quark systems, the antisymmetrizer is usually simplified as
\begin{eqnarray}
	{\cal A}&=&1-9P_{36}.
\end{eqnarray}
In Table~\ref{sfc}, we list some relevant SFC coefficients for present work. It is clear that the averaged value $\langle \mathcal{A}^{sfc} \rangle$ equals 2 for the $D_{03}$ system, which exhibits a highly strong quark change effect in this system.

Besides $D_{03}$, its mirror state $D_{30}$ has also attracted great attention from experimentalists and theorists, because according to the symmetry analysis of J. Dyson~\cite{Dyson:1964xwa}, $D_{30}$ should also have the same binding energy as $D_{03}$ has. However, till now, the newest experiment suggested that the $D_{30}$ should either be weakly bound or lie above the $\Delta \Delta$ threshold~\cite{WASA-at-COSY:2016bha}. One possible reason could be that this structure may lie near the $\Delta \Delta$ threshold and have a broad width. Nonetheless, fundamentally speaking, its weak bound or unbound characteristics should be caused by the repulsive feature of the short-range interaction of exchanged vector particles, that is, gluon and/or vector mesons, in such a spin-isospin six quark system. This can also be seen from the differences of SFC coefficients between $D_{03}$ and $D_{30}$ systems in Table~\ref{sfc}, where the operators $(\bm{\sigma}_i \cdot \bm{\sigma}_j)(\lambda^c_i \cdot \lambda^c_j)$ and $\sum\limits_{k=1}^3\lambda^F_i(k)\lambda^F_j(k)$ play important roles in the OGE and VME potentials, respectively. To verify this assertion, we calculated the properties of the $D_{30}$ state with the same method and the same sets of model parameters based on the great success in explaining the data of $D_{03}$. The obtained results are also tabulated in Table~\ref{dibaryons}. From this table, one finds that by using the models used above, the binding energy of $D_{30}$ is about $5\sim13~\rm{MeV}$, which is compatible with the newest data and is almost independent of the parameter $b_u$. In particular, the results in the extended chiral SU(3) constituent quark model with $f/g=0$ and no OGE cases agree with the experimental finding. Together with the best result for $D_{03}$, it seems that using the extended chiral SU(3) constituent quark model with $f/g=0$, the data of $D_{03}$ and $D_{30}$ can be well explained simultaneously. Once again, these findings support our previous conclusion that the VME interaction plays an important role in the short-range interactions in nonstrange systems.

\begin{figure}[H]
	\centering
	\includegraphics[scale=0.46]{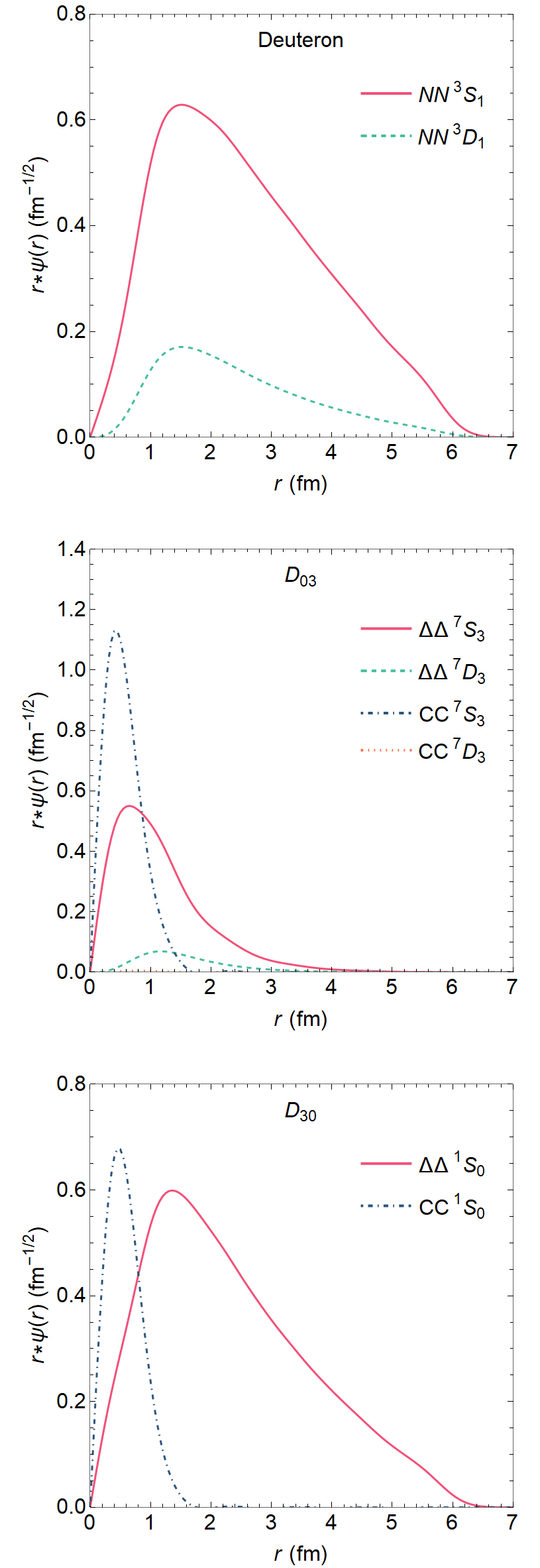}
	\vspace{0.0cm} \caption{Relative wave functions for deuteron, $D_{03}$, and $D_{30}$ in the extended chiral SU(3) quark model with $f /g=0$ and $b_u=0.45~\rm{fm}$.}
	\label{wavefunction}
\end{figure}

As a symbiotic result, we also obtain corresponding wave functions for deuteron, $D_{03}$, and $D_{30}$. We plot the relative wave functions by using the extended chiral SU(3) constituent quark model with $f /g=0$ and $b_u=0.45~\rm{fm}$ in Figure~\ref{wavefunction}. Evidently, the wave function curves of deuteron, $D_{03}$, and $D_{30}$ exhibit quite different behaviors and consequently the RMS radii. The wave function of deuteron exhibits a wide distribution for both $S-$wave and $D-$wave, so the RMS radius is quite large. The wave function of $D_{03}$ shows a very narrow and very large distribution of the CC component and a more or less wide and small distribution of the $\Delta\Delta$ component, so its RMS radius is small, indicating the formation of a compact six-quark state. The resultant values of ${\mathcal R}_{\Delta \Delta/{\rm CC}}$ and ${\mathcal R}_{6q}$ for $D_{03}$ are quite similar, which strongly suggests that $D_{03}$ is dominated by a hexaquark configuration. In contrast, although $D_{30}$ also has a narrow wave function of the CC component, it is not as large as in $D_{03}$. Meanwhile, it has a much wider wave function of the $\Delta\Delta$ component, which is larger compared with its CC component. Therefore, its RMS radius is also relatively large. The characteristics of these wave functions are consistent with the binding properties of the corresponding states obtained above. The features of these wave functions again imply that owing to the large RMS radius and lack of CC component, the deuteron can hardly be used to distinguish the relative importance between OGE and VME in short distance. On the contrary, due to existence of richer CC components, $D_{03}$ with a short range attractive feature and $D_{30}$ with a short range repulsive character can be used as excellent platforms to explore which of VME and OGE is responsible for the short range interactions.

\section{SUMMARY}\label{sec:4}
In order to reveal the dynamic mechanism of short range interactions between $u/d$ quarks, we perform a systematic analysis of the $NN$, $D_{03}$, and $D_{30}$ systems in terms of the (extended) chiral SU(3) constituent quark models. The OGE interaction and hidden color components can be easily dealt with in these constituent quark models. All model parameters are fixed by fitting the ground state masses of nonstrange baryons and binding energy of deuteron. Then, we calculate the $NN$ phase shifts and the properties of $D_{03}$ and $D_{30}$ dibaryons by using the RGM in two types of chiral SU(3) constituent quark models with different sets of model parameters and carefully examine the effects of OGE and VME interactions in the short range.

Our results show that in the extended chiral SU(3) constituent quark model, the binding energies of $D_{03}$ and $D_{30}$ are $58\sim111$ MeV and $5\sim12$ MeV, respectively. The extended chiral SU(3) constituent quark model with $f/g=0$ and $b_u=0.45~{\rm fm}$ should be adopted after further considerations are able to adequately explain all the data of the ground state masses of non-strange baryons, the $NN$ phase shifts, the binding energy of deuteron, the binding energy of $D_{03}$ dibaryon and its decay properties, and the binding behavior of $D_{30}$ dibaryon. In this case, deuteron has a binding energy of 2.25 MeV and RMS radius $\mathcal{R}_{NN}$ of 2.49 fm, respectively, while $D_{03}$ has a binding energy and a RMS radius $\mathcal{R}_{\Delta\Delta/{\rm CC}}$ of 80.08 MeV and $0.73~{\rm fm}$, and $D_{30}$ has a binding energy and a RMS radius $\mathcal{R}_{\Delta\Delta/{\rm CC}}$ of 6.00 MeV and 1.94 fm, respectively. In other words, $D_{03}$ is a compact hexaquark dominated state, and $D_{30}$ is a weakly bound state. Clearly, these values agree with the observed data. For the $D_{03}$ and $D_{30}$ states, the great difference of their binding energies is largely due to the fact that in the $D_{03}$ and $D_{30}$ states, the short-range interactions of the vector particle exchange potentials present the attractive and repulsive features, respectively. Moreover, in order to explain the available data, especially the newly observed data for $D_{03}$ and $D_{30}$ simultaneously by using the unified interaction between quarks, a relatively large contribution from the vector meson exchange seems necessary. Consequently, the contribution from one gluon exchange would be reduced.

Based on these observations, we believe that the VME interactions dominate the short range interactions between $u/d$ quarks, while a small residual OGE coupling constant $g_u$ has a value of 0.273 that is more reasonable compared with the chiral SU(3) quark model and compatible with the QCD spirit. Of course, to thoroughly understand the dynamic mechanism of short range interaction, more theoretical works and experimental measurements, such as investigating the short range behavior by using the relativistic chiral forces, are still needed.

\Acknowledgements{This work was supported by the National Natural Science Foundation of China (Grant No. 12375142), the Natural Science Foundation of Hunan Province (Grant No. 2023JJ40421), the Key Project of Hunan Provincial Education Department (Grant No. 21A0039), the Youth Talent Support Program of Hunan Normal University (Grant No. 2024QNTJ14), and the National Key Research and Development Program of China (Grant No. 2020YFA0406300). We would like to thank Lian-Rong Dai for helpful discussions.}

\InterestConflict{The authors declare that they have no conflict of interest.}



\end{multicols}
\end{document}